\documentclass[preprint,secnumarabic,amssymb,nobibnotes,aps,prd]{revtex4-1}
\usepackage{graphics}
\usepackage{graphicx}

\begin{document}

\title{Plus-minus construction leads to perfect invisibility}

\author{J.C. Nacher${}^*$}
\affiliation{Department of Complex and Intelligent Systems, Future University Hakodate, 116-2 Kamedanakano-cho, Hakodate, 041-8655, Hokkaido, Japan.}
\author{T. Ochiai}
\email[Corresponding authors: ]{nacher@fun.ac.jp, ochiai@otsuma.ac.jp}
\affiliation{School of Social Information Studies, Otsuma Women's University,
2-7-1 Karakida, Tama-shi, Tokyo 206-8540, Japan}
\date{September 29, 2010}

\begin{abstract}
Recent theoretical advances applied to metamaterials have opened new avenues to design a coating
that hides objects from electromagnetic radiation and even the sight. Here, we propose a new design of cloaking devices that creates perfect invisibility in isotropic media. A combination of positive and negative refractive indices, called {\it plus-minus} construction, is essential to achieve perfect invisibility (i.e., no time delay and total absence of reflection). Contrary to the common understanding that
between two isotropic materials having different refractive indices the electromagnetic reflection is unavoidable, our method shows that surprisingly the reflection phenomena can be completely eliminated. The invented method, different from the classical impedance matching, may also find electromagnetic applications outside of cloaking devices, wherever distortions are present arising from reflections.
\end{abstract}

\maketitle

\section{Introduction}

Rapid progress in electromagnetic and meta-material science research has made it possible to envision novel and promising applications in several areas like telecommunications,
defense, medical imaging and nano-technologies \cite{exp2,exp3,meta}. Among them, the design and construction of cloaking devices that are able to guide electromagnetic fields around any object
put inside has captured the attention of many researchers. 

In anisotropic media, Pendry {\it et al} proposed a coordinate transformation over the electromagnetic fields that allows to redirect them at will \cite{pendry2006,exp1}.
The distortion of the fields originated by the transformation media is used to generate the new values for $\epsilon$ and $\mu$, defined in anisotropic media. As a result the electromagnetic waves follow the distorted space with newly generated $\epsilon$ and $\mu$ and go around the object
and hide it from the sight.

An alternative approach was introduced by Leonhardt, where a conformal mapping was used to design cloaking devices in isotropic media \cite{leon1,leon2}.
The approach describes the propagation of light based on Hamilton's analogy between the trajectory of rays in media and the motion of particles governed by classical mechanics principles. This approach is valid when the refractive index profile $n(\vec{r})$ does not strongly change over scales comparable with the wavelength of light \cite{optics,landau}. A conformal mapping is then performed between the physical space described by a complex field $z$ and an analytical functional $w(z)$ that represents the mathematical space composed of two Riemann sheets and a branch cut which connected both sheets \cite{book1,book2}. As a result the object remains hidden on the Riemann sheets and cannot be detected by an external observer. Although the device does not generate perfect invisibility, the distortions can be made relatively small to hide objects that are large in comparison with the wavelength.

 Several recent works have extended these ideas to investigate various problems like the limitation of the electromagnetic cloak with dispersive material \cite{jiang1}, the extension of the bandwidth of electromagnetic cloaks \cite{jiang2}, the case of transformation media that re-scales space without changing the topology \cite{kildishev,schurig07} and have even been applied in a different context like acoustics \cite{chen,cummer,cummer2,review}. Another interesting application of similar concepts is called carpet cloaking \cite{liu,ergin}. A carpet mirror can be used to hide an object under it. The carpet creates a bump, therefore a distortion can be visible in the reflected image. If a carpet cloak is located on top of that bump, it can bend the light so that the distortion is eliminated and it gives the impression that the mirror is flat. Researches have extended this concept to the 3D structure and conducted promising experiments \cite{ergin}.


In a recent work, we proposed a new design of an invisibility device with isotropic materials by combining a conformal mapping and a design composed of alternating positive and negative
refraction \cite{our}. We call this type of design {\it plus-minus} construction. This design allows us to get around the Nachman theorem \cite{nach,wolf} to obtain invisibility effect.
However, in order to generate perfect cloaking, we had to fine tune 
the refractive index profile at the branch cut, which represents an
inconvenient for a practical realization of the device. In contrast, here we show that the plus-minus construction
surprisingly cancels the reflection phenomena out without
any additional modification, even when there is discontinuity of refractive
indices at any boundary, including the branch cut, which usually led to an inevitable reflection.

 In this work, we propose a novel method that uses a combination of positive and negative refraction indices, called {\it plus-minus construction}, that enables us not only to enclose the trajectory of light, but also to achieve perfect invisibility in isotropic media, without both phase delay and reflection.

In particular, here we introduce a new construction of cloaking devices by performing three successive conformal maps on the space defined by a trivial flat metric. This operation results
 on a physical space $R^2$ with a non trivial refractive index. A technical advantage of using first a flat space metric
is that the trajectory of light is trivial in the mathematical space and can be easily derived in the physical space through the conformal mappings. Hence, it is unnecessary to consider the motion equation for light rays a priori. As we show here, the trajectory is readily obtained as the result of operating three successive conformal mappings that rotate the space. Interestingly, the proposed dielectric media leads to perfect invisibility with absence of both phase delay and reflection. These findings strongly highlight the role of the negative refraction in material sciences. Although currently the application of negative refraction \cite{vesa} is confined to the construction of perfect lens \cite{pendry2000}, our results show that a device consisting of alternating positive and negative refraction ({\it plus-minus construction}) exhibits extraordinary properties like the perfect cancelation of reflection phenomena.

Our proposed design has a discontinuity at each boundary between two media with different refractive indices. It is of common knowledge that discontinuities generate reflections leading to distortions. It has been believed that reflections are inevitable in isotropic media.  Surprisingly, the invented method completely suppresses all reflections that occur at the boundary between two isotropic materials having different indices and readily leads to perfect invisibility.

The absolute control of reflection phenomena promises to bring considerable technical improvements in diverse areas, encompassing electrical engineering, telecommunications, optoelectronics and microelectronics. The practical possibility of guiding electromagnetic fields in total absence of reflection phenomena, without using the classical impedance matching, has relevance to the design and implementation of many optical and electromagnetic applications outside of cloaking devices, wherever distortions are present arising from reflections.

\section{Geometry and media}
Our theory is based on concepts of mapping between non-Euclidean geometry on
curved space, permittivity $\epsilon$ and permeability $\mu$ of materials. We briefly review
some fundamental concepts as follows \cite{pendry2006,review}:

In material, the light rays can be bent because of non-trivial permittivity and permeability tensors. However, in non-euclidian geometry, light in vacuum can also be bent by non trivial metric tensors. We can then design an invisibility device by using a mapping between a metric tensor in curved space and the characteristics $\epsilon$ and $\mu$ defined in physical space given a material.
The main idea behind this construction is that the trajectory of light in a material can be identified by that of non-Euclidean geometry.

Let $\epsilon^{ij}$ and $\mu^{ij}$ be permittivity and permeability tensors and $\gamma_{ij}$ be the metric tensor in the physical space. And let $g_{ij}$ be the metric tensor in the mathematical space. The connection between the metric $g_{ij}$ in the mathematical space and both permittivity $\epsilon^{ij}$ and permeability $\mu^{ij}$ tensors in physical space is given by
\begin{eqnarray}
\epsilon^{ij}=\mu^{ij}=\frac{\sqrt{g}}{\sqrt{\gamma}}g^{ij}
\end{eqnarray}
where $g=det(g_{ij})$ is the determinant of $g_{ij}$, and $\gamma=det(\gamma_{ij})$ is the determinant of $\gamma_{ij}$. Namely, $\sqrt{g}$ is the volume element of mathematical space, and $\sqrt{\gamma}$ is the volume element of physical space.

\section{Isotropic media}
In this paper, we focus our attention on isotropic media.

Since, in isotropic case, $\epsilon^{ij}$ and $\mu^{ij}$ tensors are proportional to the identity matrix,  we can set
\begin{eqnarray}\label{eqn:isotropic metric tensor}
g_{ij}=n^2\delta_{ij}
\end{eqnarray}
where $\delta_{ij}$ is Kronecker delta function, and $n$ is an arbitrary scalar function which can be interpreted as the refractive index because of the following reason.
In isotropic media, the metric tensor is given by
\begin{eqnarray}
ds^2=g_{ij}dx^idx^j=n^2(dx^2+dy^2+dz^2)
\end{eqnarray}
therefore the line element is written as
\begin{eqnarray}
ds=n\sqrt{(dx^2+dy^2+dz^2)}
\end{eqnarray}
From this, it is clear that the $n$ should be interpreted as a refractive index.

\section{Conformal mapping}
Conformal mappings keep the metric form proportional to the identity as shown in (\ref{eqn:isotropic metric tensor}). In this section, we discuss the transformation law under the conformal mappings.
Since complex analytic function is a conformal mapping, we will consider analytic functions on $C$ in our analysis. For $z=x+iy$, we consider the analytic function as follows:
\begin{eqnarray}
z^\prime=f(z)=x^\prime(x,y)+iy^\prime(x,y)
\end{eqnarray}
where $i^2=-1$.
Then, the Cauchy Riemann equation holds
\begin{eqnarray}
&&\frac{\partial x^\prime}{\partial x}=\frac{\partial y^\prime}{\partial y}\\
&&\frac{\partial x^\prime}{\partial y}=-\frac{\partial y^\prime}{\partial x}
\end{eqnarray}
On $z^\prime=x^\prime+iy^\prime$ space, let us set the metric for isotropic media
\begin{eqnarray}
ds^2=n^{\prime 2}((dx^{\prime})^2+(dy^{\prime})^2)
\end{eqnarray}

For conformal mapping, the metric is changed as follows.
\begin{eqnarray}
ds^2=n^{\prime 2}\Bigl((\frac{\partial x^\prime}{\partial x})^2+(\frac{\partial x^\prime}{\partial y})^2\Bigr)(dx^2+dy^2)
\end{eqnarray}
Therefore, the transformation law of refractive index between $z$ space and $z^\prime$ space is given by
\begin{eqnarray}\label{eqn: the transformation law}
n^2=n^{\prime 2}\Bigl((\frac{\partial x^\prime}{\partial x})^2+(\frac{\partial x^\prime}{\partial y})^2\Bigr)
\end{eqnarray}
We note that this expression can be shown to be the same as the expression in \cite{leon2} by proper variable transformation.

\section{Cloaking device construction and design}
We construct our invisibility device by using two ingredients.

\begin{enumerate}
\item
First, we operate three conformal mappings which transform the flat space into a curved space with non trivial metric.
\item
Next, the negative refraction property of meta-materials which, creates a mirror-like image, enables us to enclose the trajectory of light.
\end{enumerate}

\subsection{Conformal map}
We perform three conformal mappings in order to construct our invisibility device.

Let $Z$ be a complex plane and be identified with physical space. The first conformal mapping $f_1:Z\to W_1$ is given by
\begin{eqnarray}\label{eqn:conformal map f1}
w_1=f_1(z)=z+\frac{a^2}{z}
\end{eqnarray}
where $z\in Z$, $w_1\in W_1$. Here $W_1$ is a Riemann surface defined by $f_1$.
$W_1$ space consists of two sheets, the first and second Riemann sheets ($W_{1f}$ and $W_{1s}$).
By the conformal map (\ref{eqn:conformal map f1}), the outside of the circle ($|z|>=a$) is mapped to the first Riemann sheet $W_{1f}$, and the inside of the circle ($|z|<a$) is mapped to the second Riemann sheet $W_{1s}$. Hence, $W_1=W_{1f} \cup W_{1s}$. The branch cut is given by $-2a<Re(w_1)<2a$, $Im(w_1)=0$, which is mapped to $|z|=a$ in $Z$ space. This conformal map is first introduced for invisibility devices in \cite{leon1}.

For the second Riemann sheet $W_{1s}$, we define the two conformal maps as follows. The conformal map $f_2:W_{1s}\to W_2$ is
\begin{eqnarray}
&&w_2=f_2(w_1)=-2a+4a(\frac{w_1+2a}{4a})^{\frac{1}{3}}\label{eqn:conformal map f2}
\end{eqnarray}
where $w_1\in W_{1s}$ and $w_2 \in W_2$.
The conformal map $f_3:W_2\to W_3$ is
\begin{eqnarray}
&&w_3=f_3(w_2)=2a-4a(\frac{-w_2+2a}{4a})^{3}\label{eqn:conformal map f3}
\end{eqnarray}
where $w_2\in W_2$ and $w_3\in W_3$.  The domain and region of map $f_2$ and $f_3$ will be given in the next section, in order to make $f_2$ and $f_3$ bijective and restrict both the domain and range of $f_2$ and $f_3$ to the inside the upper part of $W_{1s}$, $W_2$ and $W_3$, respectively. Here, we remark that $(2a,0)$ and $(-2a,0)$ are the fixed points (invariant points) under both mappings $f_2$ and $f_3$.

On the other hand, we do not perform any transformation on the first Riemann sheet $W_{1f}$, so it is kept invariant. We will discuss later
the problem of discontinuity of this mapping on the branch cut $-2a<Re(w_1)<2a$, $Im(w_1)=0$ between $W_{1f}$ and $W_{1s}$.

We summarize the maps $f_1$, $f_2$ and $f_3$ as follows.
\begin{eqnarray}
&&Z \stackrel{f_1}{\to} W_1=W_{1f} \cup W_{1s}\\
&&W_{1s} \stackrel{f_2}{\to} W_2 \stackrel{f_2}{\to} W_3
\end{eqnarray}

The inverse function of $f_1$, $f_2$ and $f_3$ are then given by:
\begin{eqnarray}
&&z=f_1^{-1}(w_1)=\frac{1}{2}(w_1\pm\sqrt{w_1^2-4a^2})\label{eqn:conformal map inverse f1}\\
&&w_1=f_2^{-1}(w_2)=-2a+4a(\frac{w_2+2a}{4a})^{3}\label{eqn:conformal map inverse f2}\\
&&w_2=f_3^{-1}(w_3)=2a-4a(\frac{-w_3+2a}{4a})^{\frac{1}{3}}\label{eqn:conformal map inverse f3}
\end{eqnarray}

We will successively perform three conformal mappings $f_3^{-1}$, $f_2^{-1}$, $f_1^{-1}$ to construct the invisibility device.

\begin{figure}
\includegraphics[scale=0.3]{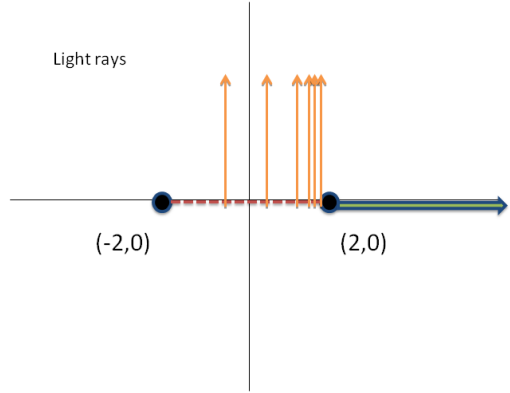}
\caption{\label{fig:rot image1} Representation of light trajectory in $W_3$ space. ($a=1$) Orange lines are light trajectory in $W_3$ space. By the mapping $f_3^{-1}$, the arrow in Fig. \ref{fig:rot image1} rotates 120 degree around the right hand side of branch cut $(2a,0)$, which is shown in Fig. \ref{fig:rot image2}}
\end{figure}

\begin{figure}
\includegraphics[scale=0.3]{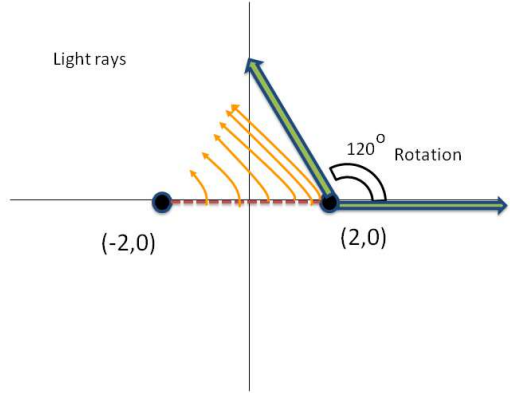}
\caption{\label{fig:rot image2} Illustration of light trajectory in $W_2$ space. By the mapping $f_3^{-1}$ the arrow in Fig. \ref{fig:rot image1} rotates 120 degree. Moreover, by the mapping $f_3^{-1}$, the light trajectories in Fig. \ref{fig:rot image1} are bent to those shown in  Fig. \ref{fig:rot image2}.}
\end{figure}

\begin{figure}
\includegraphics[scale=0.3]{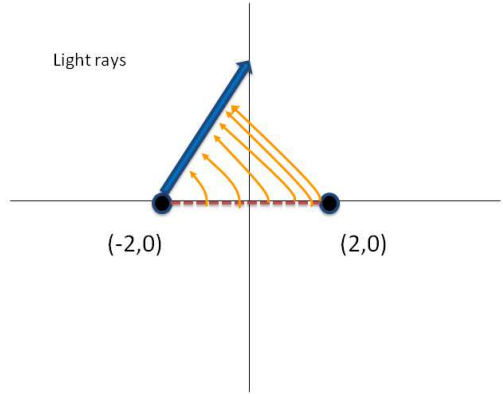}
\caption{\label{fig:rot image3} Image of light trajectory in $W_2$ space, which is the same as in  Fig. \ref{fig:rot image2}, but the location of the arrow is different. ($a=1$) Orange lines are light trajectory in $W_2$ space. By the mapping $f_2^{-1}$, the arrow in Fig. \ref{fig:rot image3} rotates 120 degree more around the left hand side of branch cut $(-2a,0)$, which is shown in Fig. \ref{fig:rot image4}.}
\end{figure}

\begin{figure}
\includegraphics[scale=0.3]{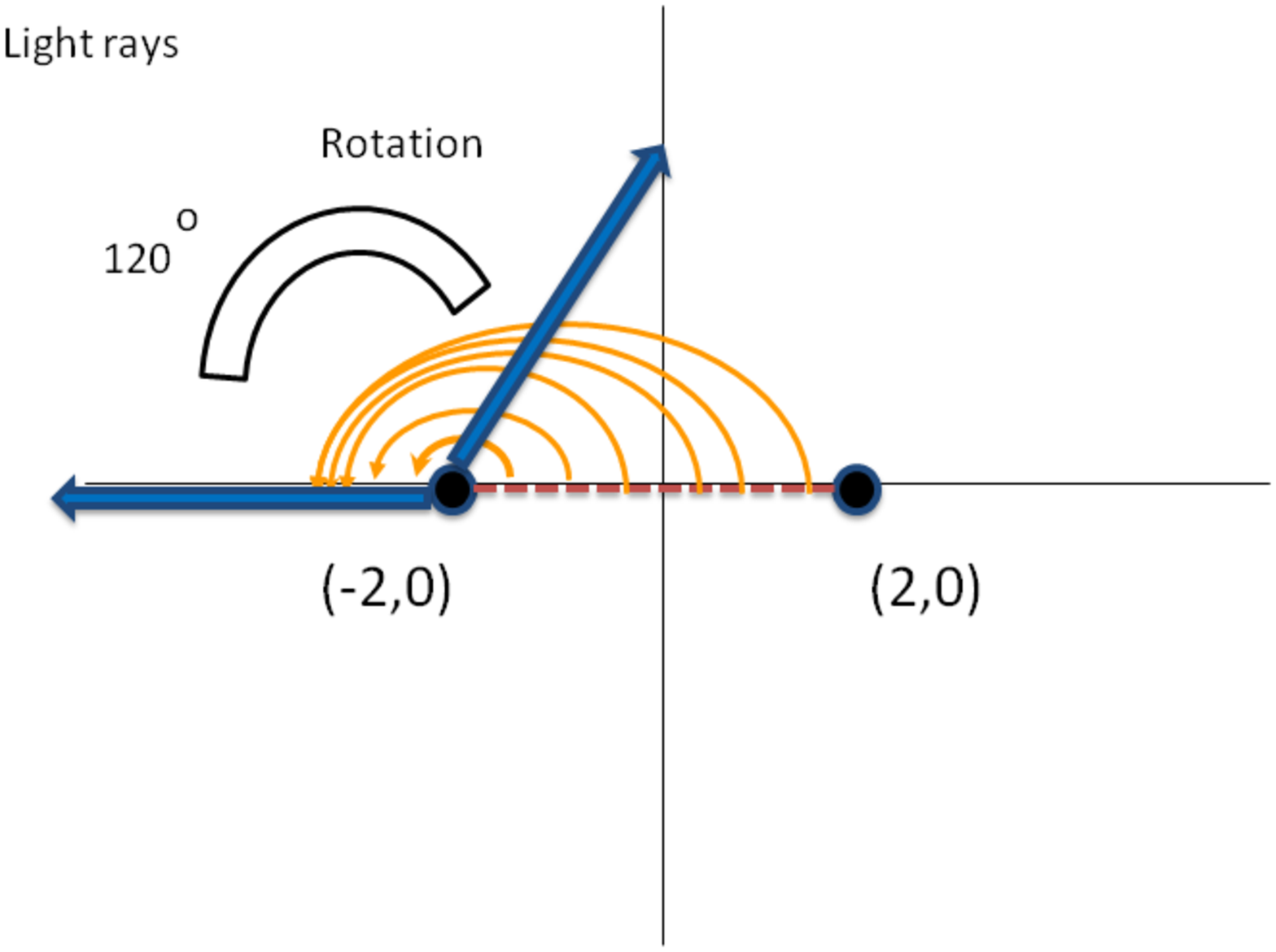}
\caption{\label{fig:rot image4}Representation of light trajectory in $W_1$ space. By the mapping $f_2^{-1}$ the arrow in Fig. \ref{fig:rot image3} rotates 120 degree more. Moreover, by the mapping $f_3^{-1}$, the light trajectories in Fig. \ref{fig:rot image3} are bent more to those shown in  Fig. \ref{fig:rot image4}. }
\end{figure}

We show an illustration of the mapping $f_3^{-1}$ in Fig. \ref{fig:rot image1} and Fig. \ref{fig:rot image2}. The mapping $f_2^{-1}$ can be visualized in Fig. \ref{fig:rot image3} and Fig. \ref{fig:rot image4}.

%

%

\subsection{Domain and Region of $f_2$ and $f_3$}
In this section, we set the domain and region of $f_2$ and $f_3$, defined in the previous section.

\subsubsection{Domain and region in $W_2$}
We set the boundary of both the domain of $f_3$ and the region of $f_2$ in $W_2$ as follows:
\begin{eqnarray}
I_1=\{re^{i\theta}-2a | 0<r<4a, \theta=\frac{1}{3}\pi \} \label{eqn: I1} \\
I_2=\{re^{i\theta}+2a | 0<r<4a, \theta=\frac{2}{3}\pi \} \label{eqn: I2}
\end{eqnarray}
The triangle region enclosed by $I_1$, $I_2$ and $Im(w_2)=0$ is the domain of $f_3$ and the region of $f_2$. $I_1$ and $I_2$ are shown in Fig. \ref{fig: w2}. In what follows, this region is denoted by $I$.

\subsubsection{Region in $W_3$}
We set the boundary of the region of $f_3$ in $W_3$ as follows:
\begin{eqnarray}
&&J_1=f_3(I_1)\label{eqn: J1}\\
&&J_2=f_3(I_2)=\{x+iy | 2a<x<6a, y=0\}\label{eqn: J2}
\end{eqnarray}
The region enclosed by $J_1$, $J_2$ and $Im(w_3)=0$ is the region of $f_3$.
$J_1$ and $J_2$ are shown in Fig. \ref{fig: w3}. In what follows, this region is denoted by $J$.

\subsubsection{Domain in $W_1$}
We set the boundary of the domain of $f_2$ in $W_{1s}$ as follows:
\begin{eqnarray}
&&K_1=f_2^{-1}(I_1)=\{x+iy | -6a<x<-2a, y=0\}\label{eqn: K1}\\
&&K_2=f_2^{-1}(I_2)\label{eqn: K2}
\end{eqnarray}
The region enclosed by $K_1$, $K_2$ and $Im(w_1)=0$ is the domain of $f_2$.
$K_1$ and $K_2$ are shown in Fig.  \ref{fig: w1}. In what follows, this region is denoted by $K$.

All orange (continuous), green (dash-dotted) and blue (dotted) lines in Figs. \ref{fig: w3}, \ref{fig: w2}, \ref{fig: w1}, \ref{fig: z}, correspond each other by the three conformal maps $f_1$, $f_2$ and $f_3$.

\subsection{Refractive index}
In this section, we will compute the refractive index on physical space $Z$, by applying the transformation law of conformal mappings (\ref{eqn: the transformation law}) and combining negative refractive index as mirror-like image.

\subsubsection{Polar coordinate}
We set the coordinate in each complex space $Z$, $W_1$, $W_2$, $W_3$ space as follows. For each space, we set
\begin{eqnarray}
&&z=x+iy~~~~(z\in Z)\\
&&w_1=x_1+iy_1~~~~(w_1\in W_1=W_{1f}\cup W_{1s})\\
&&w_2=x_2+iy_2~~~~(w_2\in W_2)\\
&&w_3=x_3+iy_3~~~~(w_3 \in W_3)
\end{eqnarray}

We introduce a polar coordinate in $W_3$ space as follows.
\begin{eqnarray}
&&x_3=2a+r_3\cos\theta_3\\
&&y_3=r_3\sin\theta_3
\end{eqnarray}
Here we note that the origin of the polar coordinate is located at the right hand side point of the branch cut $(2a,0)$.

We introduce a polar coordinate in $W_2$ space in two ways as follows.
\begin{eqnarray}
&&x_2=2a+r_2\cos\theta_2=-2a+r_2^\prime\cos\theta_2^\prime\\
&&y_2=r_2\sin\theta_2=r_2^\prime\sin\theta_2^\prime
\end{eqnarray}
Here we note that the origin of the polar coordinate $(r_2, \theta_2)$ is located at the right hand side point of the branch cut $(2a,0)$. Similarly, the origin of the polar coordinate $(r_2^\prime, \theta_2^\prime)$ is located at the left hand side point of the branch cut $(-2a,0)$.

Therefore, using (\ref{eqn:conformal map f3}), we obtain
\begin{eqnarray}
&&\frac{r_3}{4a}=(\frac{r_2}{4a})^3\\
&&\theta_3=3\theta_2
\end{eqnarray}

We introduce a polar coordinate in $W_1=W_{1f}\cup W_{1s}$ space as follows.
\begin{eqnarray}
&&x_1=-2a+r_1\cos\theta_1\\
&&y_1=r_1\sin\theta_1
\end{eqnarray}
Here we note that the origin of the polar coordinate is located at the left hand side point of the branch cut $(-2a,0)$.

Therefore, using (\ref{eqn:conformal map f2}), we obtain
\begin{eqnarray}
&&\frac{r_1}{4a}=(\frac{r_2^\prime}{4a})^3\\
&&\theta_1=3\theta_2^\prime
\end{eqnarray}

We introduce a polar coordinate in $Z$ space as follows.
\begin{eqnarray}
&&x=r\cos\theta\\
&&y=r\sin\theta
\end{eqnarray}

In the next section, we compute the refractive index on the upper area $J$, $I$, $K$ in $W_3$, $W_2$ and $W_{1s}$, and set the refractive index of symmetric area of $J$, $I$, $K$ with respect to real axis as mirror-like image, respectively, but the sign of refractive index is opposite.

\subsubsection{Refractive index on $W_3$}
For the region $J$ in the upper area of $W_3$ (the area surrounded by real axis, $J_1$ and $J_2$ (See (\ref{eqn: J1}) and (\ref{eqn: J2})), we set the flat metric as follows
\begin{eqnarray}
ds^2&=&(dx_3)^2+(dy_3)^2\nonumber\\
&=&(dr_3)^2+(r_3)^2(d\theta_3)^2
\end{eqnarray}
Considering this, we can set the refractive index in $J$ as follows
\begin{eqnarray}
n_3=1
\end{eqnarray}
Therefore in $J$ area (in $J \subset W_3$), all the light ray trajectories go on the straight line, since the metric is flat. (See Fig. \ref{fig: w3})

On the other hand, in the symmetric region of $J$ with respect to real axis in  $W_3$, we set the refractive index like a mirror like image but the sign of refractive index is opposite
\begin{eqnarray}
n_3=-1
\end{eqnarray}
Therefore, in the symmetric region of $J$ all the light ray trajectories also go on the straight line, since the metric is flat too. However the phase velocity is opposite to $n_3=1$ case (See Fig. \ref{fig: w3}).

\subsubsection{Refractive index on $W_2$}

By applying the transformation law for the conformal map (\ref{eqn: the transformation law}) to $f_3$, we can compute the refractive index $n_2$ on $I$, which is the domain of $f_3$ and the region of $f_2$ in $W_2$. In the area $I$, the induced metric is given by
\begin{eqnarray}
ds^2=(n_2)^2\{(dr_2)^2+(r_2)^2(d\theta_2)^2\}
\end{eqnarray}
where
\begin{eqnarray}
n_2^2=9(\frac{r_2}{4a})^4
\end{eqnarray}
Therefore in $I$ region, the refractive index is given by
\begin{eqnarray}
n_2=3(\frac{r_2}{4a})^2
\end{eqnarray}
And in the symmetric region of $I$ with respect to real axis in  $W_2$, the refractive index is
\begin{eqnarray}
n_2=-3(\frac{r_2}{4a})^2
\end{eqnarray}
(See Fig. \ref{fig: w2}).

\subsubsection{Refractive index on $W_{1s}$}
In the same way, we apply the transformation law for conformal map (\ref{eqn: the transformation law}) to $f_2$, we obtain the refractive index $n_1$ in $K$ which is  the domain of $f_2$ in $W_{1s}$. In the $K$, the induced metric is given by
\begin{eqnarray}
ds^2=n_1^2\{(dr_1)^2+(r_1)^2(d\theta_1)^2\}
\end{eqnarray}
where
\begin{eqnarray}
n_1^2=(\frac{r_2}{4a})^4(\frac{r_1}{4a})^{-\frac{4}{3}}
\end{eqnarray}
Therefore in $K$ region, the refractive index is given by
\begin{eqnarray}
n_1=(\frac{r_2}{4a})^2(\frac{r_1}{4a})^{-\frac{2}{3}}
\end{eqnarray}
And in the symmetric region of $K$ with respect to real axis in  $W_{1s}$, the refractive index is
\begin{eqnarray}
n_1=-(\frac{r_2}{4a})^2(\frac{r_1}{4a})^{-\frac{2}{3}}
\end{eqnarray}
(See Fig. \ref{fig: w1}).

\subsubsection{Refractive index inside the circle $|z|<a$ in Z}
Finally, we apply a transformation law for conformal map (\ref{eqn: the transformation law}) to $f_1$. On the region $R_1=f_1^{-1}(K)=\{r e^{i\theta}\in Z | r<a, -\pi\leq \theta <0\}$ in physical space $Z$, the induced metric is given by
\begin{eqnarray}
ds^2=n^2(dr^2+r^2d\theta^2)
\end{eqnarray}
where
\begin{eqnarray}
n^2=(\frac{r_2}{4a})^4(\frac{r_1}{4a})^{-\frac{4}{3}}(1-\frac{2a^2}{r^2}\cos 2\theta+\frac{a^4}{r^4})
\end{eqnarray}

Therefore we obtain the refractive index $n$ on the region $R_1$ in physical space $Z$ as follows,
\begin{eqnarray}
n=\sqrt{(\frac{r_2}{4a})^4(\frac{r_1}{4a})^{-\frac{4}{3}}(1-\frac{2a^2}{r^2}\cos 2\theta+\frac{a^4}{r^4})}
\end{eqnarray}
On the other hand, the refractive index $n$ on the region $R_2=\{r e^{i\theta}\in Z | r<a, 0\leq \theta <\pi\}$ in physical space $Z$ as follows,
\begin{eqnarray}
n=-\sqrt{(\frac{r_2}{4a})^4(\frac{r_1}{4a})^{-\frac{4}{3}}(1-\frac{2a^2}{r^2}\cos 2\theta+\frac{a^4}{r^4})}
\end{eqnarray}

\subsubsection{Refractive index outside the circle $|z|\geq a$ in Z}
Finally, We set flat metric on the first Riemann sheet of $W_{1f}$ as follows:
\begin{eqnarray}
ds^2&=&(dx_1)^2+(dy_1)^2\nonumber\\
&=&(dr_1)^2+(r_1)^2(d\theta_1)^2
\end{eqnarray}
Therefore, in  all $W_{1f}$ region, the refractive index is given by
\begin{eqnarray}
n_1=1
\end{eqnarray}
Hence, all the light rays go on the straight lines in $W_{1f}$.

We apply a transformation law (\ref{eqn: the transformation law}) to $f_1$. Then, for the region $R_3=\{r e^{i\theta}\in Z | r\geq a\}$, we obtain
\begin{eqnarray}
n^2=(1-\frac{2a^2}{r^2}\cos 2\theta+\frac{a^4}{r^4})~~~~(r\geq a)
\end{eqnarray}
Thus, the refractive index $n$ on the region $R_3$ in physical space $Z$ is given by
\begin{eqnarray}
n=\sqrt{(1-\frac{2a^2}{r^2}\cos 2\theta+\frac{a^4}{r^4})}~~~~(r\geq a)
\end{eqnarray}

\subsubsection{Summary of Refractive index in Z}
Then, we summarize the refractive index on all area of physical space.
\begin{eqnarray}
n=\left\{
\begin{array}{cc}\label{eqn: summary of refractive index}
\sqrt{(1-\frac{2a^2}{r^2}\cos 2\theta+\frac{a^4}{r^4})} & (r \geq a)\\
\sqrt{\frac{(r_2/4a)^4}{(r_1/4a)^{4/3}}(1-\frac{2a^2}{r^2}\cos 2\theta+\frac{a^4}{r^4})} & (r<a ~~\mbox{and} -\pi\leq \theta <0)\\
-\sqrt{\frac{(r_2/4a)^4}{(r_1/4a)^{4/3}}(1-\frac{2a^2}{r^2}\cos 2\theta+\frac{a^4}{r^4})} & (r<a ~~\mbox{and}~~ 0\leq \theta <\pi)\\
\end{array}
\right.
\end{eqnarray}
where
\begin{eqnarray}
&&z=re^{i\theta}\\
&&r_1=|f_1(z)+2a|\\
&&r_2=|f_2\circ f_1(z)-2a|
\end{eqnarray}

\begin{figure}
\includegraphics[scale=0.3]{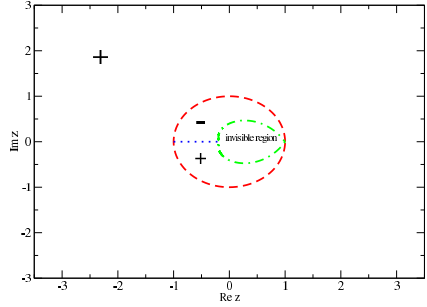}
\caption{\label{fig:plus-minus} The figure shows the positive and negative regions of refractive index in physical space $Z$ ({\it Plus-minus construction}). "$+$" (resp. "$-$") means positive (resp. negative) refractive index.}
\end{figure}

Therefore, the refractive index in physical space are mixing of positive and negative refractive index. We call this type of construction "{\it plus-minus construction}"
See Fig. \ref{fig:plus-minus}.

\section{Trajectory of light}
In this section, we compute the trajectory of light.

\begin{figure}
\includegraphics[scale=0.3]{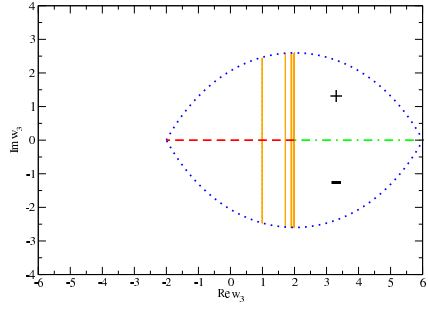}
\caption{\label{fig: w3} Orange (continuous) line represents trajectory of light in $W_3$. Red (dashed) line indicates the branch cut. "$+$" (resp. "$-$") means positive (resp. negative) refractive index. All the light go along straight lines, since the refractive index on this region is $n_3=1$ or $n_3=-1$. Blue (doted) line is given by $J_1$ in Eq. (\ref{eqn: J1}). Green (dash-dotted) line is given by $J_2$ in Eq. (\ref{eqn: J2}). }
\end{figure}

\begin{figure}
\includegraphics[scale=0.3]{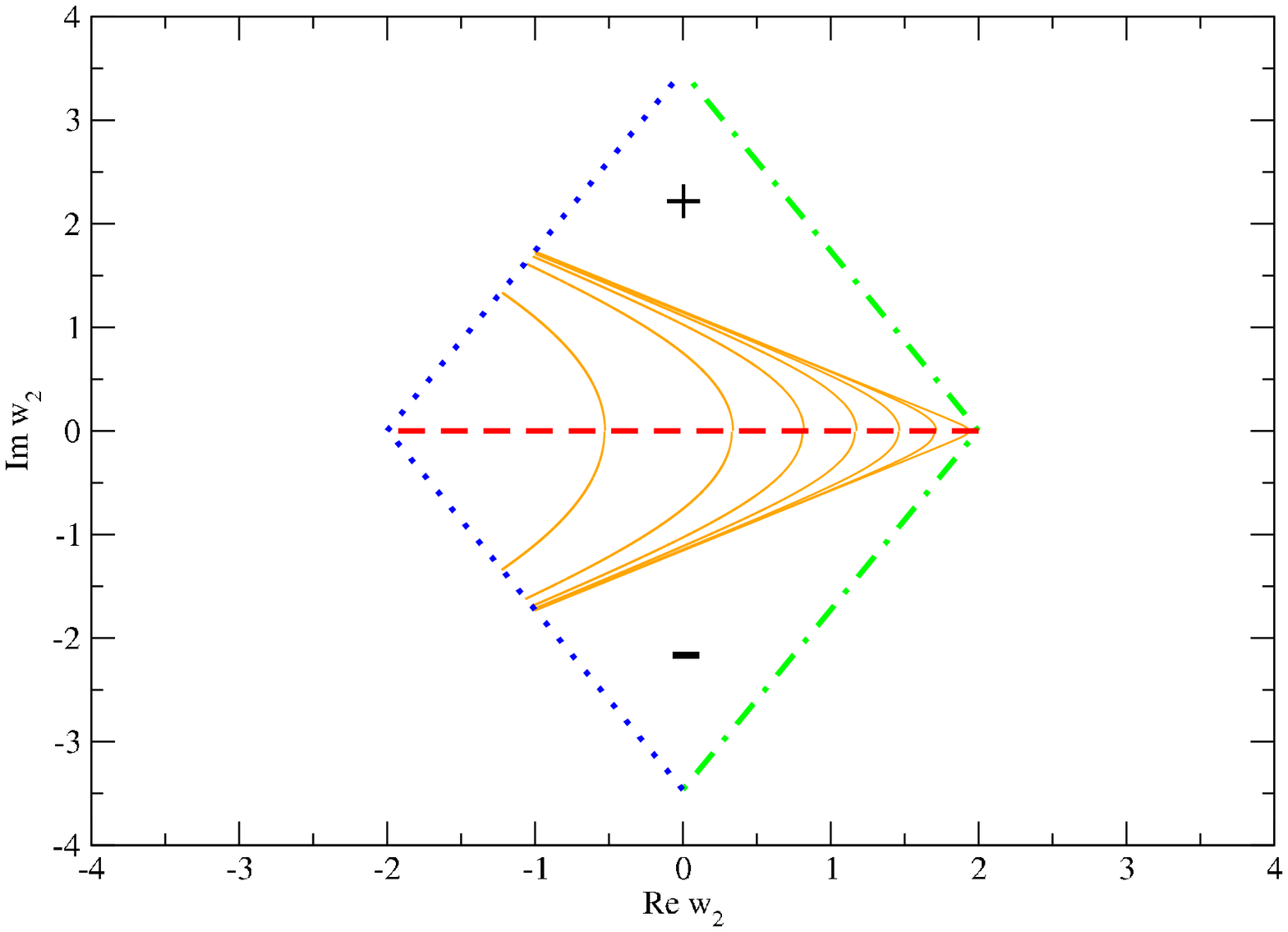}
\caption{\label{fig: w2} Trajectory of light in $W_2$. Orange (continuous) line is trajectory of light in $W_2$. "$+$" (resp. "$-$") means positive (resp. negative) refractive index. Blue (dotted) line is given by $I_1$ in Eq. (\ref{eqn: I1}). Green (dash-dotted) line is given by $I_2$ in Eq. (\ref{eqn: I2}). All the orange, blue and green lines  between Fig. \ref{fig: w2} and Fig. \ref{fig: w3} correspond to each other by conformal map $f_3^{-1}$ (see (\ref{eqn:conformal map inverse f1}))}
\end{figure}

\begin{figure}
\includegraphics[scale=0.30]{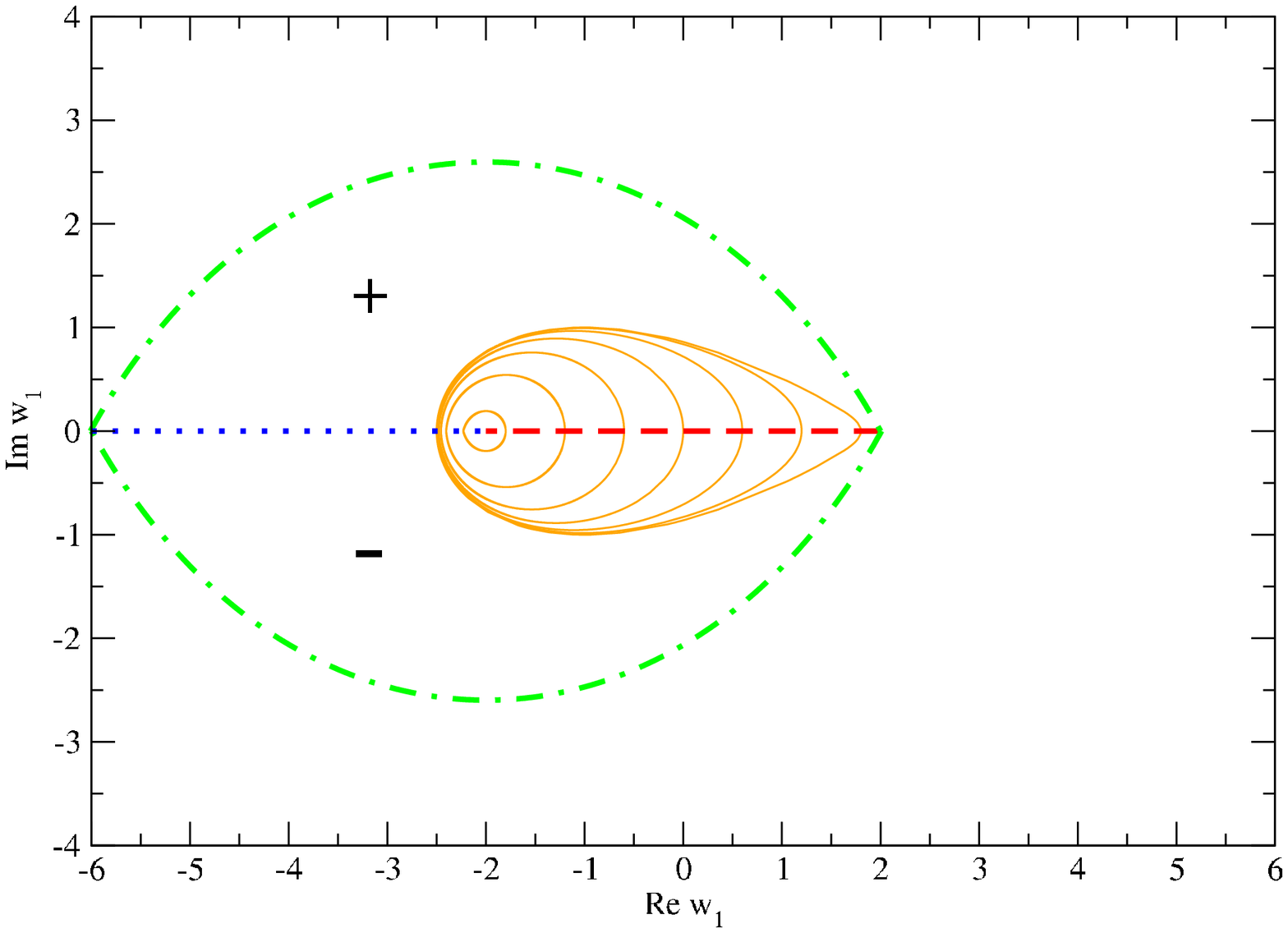}
\caption{\label{fig: w1} Trajectory of light in $W_{1s}$. Orange (continuous) line is trajectory of light in $W_1$. "$+$" (resp. "$-$") means positive (resp. negative) refractive index. Blue (dotted) line is given by $K_1$ in Eq. (\ref{eqn: I1}). Green (dash-dotted) line is given by $K_2$ in Eq. (\ref{eqn: I2}). All the orange, blue and green lines  between Fig. \ref{fig: w1} and Fig. \ref{fig: w2} correspond to each other by conformal map $f_2^{-1}$ (see (\ref{eqn:conformal map inverse f2}))}
\end{figure}

\begin{figure}
\includegraphics[scale=0.3]{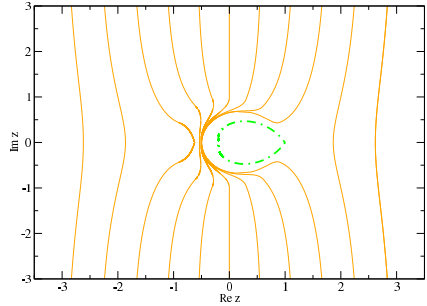}
\caption{\label{fig: z} Trajectory of light in physical space $z$. Orange (continuous) line is trajectory of light in $Z$. Green (dash-dotted) curve is given by $L$ in Eq. (\ref{eqn: L}).
The area enclosed by the green curve is invisible. All the orange, green lines  between Fig. \ref{fig: z} and Fig. \ref{fig: w1} correspond to each other by conformal map $f_1^{-1}$ (see (\ref{eqn:conformal map inverse f1}))}
\end{figure}

The trajectory of light can be obtained by successive application of the three conformal maps $f_3^{-1}$, $f_2^{-1}$ $f_1^{-1}$ from $W_3$ to $Z$.
In $W_3$, all the light rays go along straight lines, since the metric of this space is flat. In other words, the refractive index of $W_3$ is $1$ or $-1$ (see Fig. \ref{fig: w3}). As shown in Fig. \ref{fig: w2}, in $W_2$, the light rays are bent by the conformal map $f_3^{-1}$ from $W_3$. In $w_1\in W_{1s}$, the light rays
are bent more by the conformal map $f_2^{-1}$ from $W_2$ (see Fig. \ref{fig: w1}). Furthermore, all the light trajectories are enclosed by the negative refraction like a mirror-like image.
The last trajectory of light is given by applying the conformal map $f_1^{-1}$ as shown in Fig. \ref{fig: z}.

It is worth noticing that, in this example, we only show the perpendicular incident case to branch cut for the matter of convenience. However, we point out that the proposed device is invisible for any arbitrary angle of incident light rays.

On the other hand, the boundary between the device and the invisible region in physical space $Z$ can be obtained as follows:
\begin{eqnarray}
L=f_1^{-1}(K_2)\label{eqn: L}
\end{eqnarray}
Therefore, the area enclosed by $L$ and the symmetric curve of $L$ with respect to the real axis is invisible. We can hide any object inside this region as shown in Fig. \ref{fig: z}.
All the boundary represented by red, green, blue lines correspond to each other by the three conformal maps $f_3^{-1}$, $f_2^{-1}$ $f_1^{-1}$ in Figs. \ref{fig: w3}, \ref{fig: w2}, \ref{fig: w1} and \ref{fig: z}.

\newpage

\section{Perfectness of the invisibility devices}
The time delay and reflection causes the imperfectness of invisibility devices and many other electromagnetic applications. Especially, in our construction, there are several discontinuous boundaries of refractive indices which may cause reflection (see Eq. (\ref{eqn: summary of refractive index})). In spite of that,  we will show that our device does not generate phase delay and reflection and leads to perfect invisibility.

\subsection{No phase delay}
Negative refractive index has opposite sign of phase delay to the positive refractive index. The phase delay in the positive refractive index region $K$ in the second Riemann sheet of $W_{1s}$ is canceled by the phase shift in the negative refractive index of the mirror like image region of $K$ (see Fig. \ref{fig: w1}). Therefore, the proposed cloaking device has no time delay because of the property of negative refractive index.

\subsection{Boundary reflection}
Discontinuities of refractive index at the boundary between two materials cause the reflection, which may distort the images and precludes the perfect invisibility.
In our device, discontinuity of refractive index exists in the following two boundaries.

\begin{enumerate}
\item
Discontinuity of refractive index at branch cut (see red (dashed) line in Fig. \ref{fig: w1}).
\item
Discontinuity of refractive index at the boundary between positive and negative refractive indices (see blue (dotted) line in Fig. \ref{fig: w1}).
\end{enumerate}

We will show that in both cases, the reflection completely vanishes, using the
striking features of the plus-minus construction.


\subsubsection{Discontinuity of refractive index at branch cut}

Let us consider the boundary between two different material 1 and 2. Let $\epsilon_1$ and $\mu_1$ (resp. $\epsilon_2$ and $\mu_2$) be permittivity and permeability of material 1 (resp. material 2). Then, the refractive index of material 1 (resp. material 2) is given by $\hat{n}_1=\sqrt{\epsilon_1\mu_1}$ (resp. $\hat{n}_2=\sqrt{\epsilon_2\mu_2}$).

By considering the standard electromagnetic theory \cite{book3}, the reflection and transmission ratio are given as follows:

\begin{enumerate}
\item
Perpendicular to plane of incidence:

The transmission ratio from material 1 to material 2 (perpendicular to plane of incidence) is given by
\begin{eqnarray}
&&\frac{E_t}{E_i}=\frac{2\sqrt{\frac{\mu_2}{\epsilon_2}}\cos i}{\sqrt{\frac{\mu_2}{\epsilon_2}}\cos i+\sqrt{\frac{\mu_1}{\epsilon_1}}\sqrt{1-(\frac{\hat{n}_1}{\hat{n}_2})^2\sin^2 i}}
\end{eqnarray}
where $i$ denotes the incident angle. Similarly, the reflection ratio is given by
\begin{eqnarray}
&&\frac{E_r}{E_i}=\frac{\sqrt{\frac{\mu_2}{\epsilon_2}}\cos i-\sqrt{\frac{\mu_1}{\epsilon_1}}\sqrt{1-(\frac{\hat{n}_1}{\hat{n}_2})^2\sin^2 i}}{\sqrt{\frac{\mu_2}{\epsilon_2}}\cos i+\sqrt{\frac{\mu_1}{\epsilon_1}}\sqrt{1-(\frac{\hat{n}_1}{\hat{n}_2})^2\sin^2 i}}\label{eqn:Reflection 1}
\end{eqnarray}

\item
Parallel to plane of incidence:

The transmission ratio from material 1 to material 2 (parallel to plane of incidence) is given by

\begin{eqnarray}
&&\frac{E_t}{E_i}=\frac{2\sqrt{\frac{\mu_2}{\epsilon_2}}\cos i}{\sqrt{\frac{\mu_1}{\epsilon_1}}\cos i+\sqrt{\frac{\mu_2}{\epsilon_2}}\sqrt{1-(\frac{\hat{n}_1}{\hat{n}_2})^2\sin^2 i}}
\end{eqnarray}

Similarly, the reflection ratio is given by
\begin{eqnarray}
&&\frac{E_r}{E_i}=\frac{\sqrt{\frac{\mu_1}{\epsilon_1}}\cos i-\sqrt{\frac{\mu_2}{\epsilon_2}}\sqrt{1-(\frac{\hat{n}_1}{\hat{n}_2})^2\sin^2 i}}{\sqrt{\frac{\mu_1}{\epsilon_1}}\cos i+\sqrt{\frac{\mu_2}{\epsilon_2}}\sqrt{1-(\frac{\hat{n}_1}{\hat{n}_2})^2\sin^2 i}}\label{eqn:Reflection 2}
\end{eqnarray}

\end{enumerate}

\begin{figure}
\includegraphics[scale=0.3]{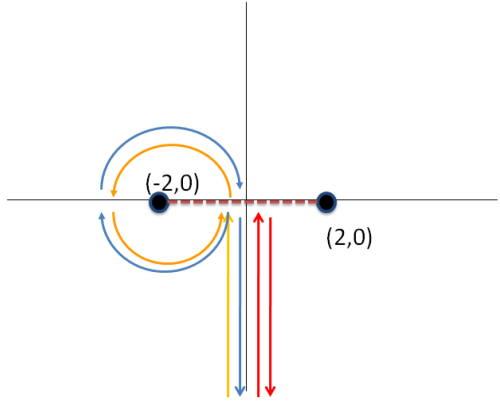}
\caption{\label{fig:reflection image} Light rays trajectory in $W_1$ space. Red arrows are reflected light rays at the branch cut. Orange and blue arrows show another reflected ray lights. Orange arrows indicate the light rays that enter from first Riemann sheet $W_{1f}$ to the second Riemann sheet $W_{1s}$ through the first branch cut. After that these light rays can also be reflected at the second branch cut. This reflection is shown by blue arrows. Blue arrows show that the light rays return to the first Riemann sheet $W_{1f}$. Although figure shows perpendicular incident light rays to the branch cut, the cancelation of the reflection is valid for any arbitrary incidence angle.}
\end{figure}

If we consider the impedance matching case,
\begin{eqnarray}
&&\sqrt{\frac{\mu_1}{\epsilon_1}}=\sqrt{\frac{\mu_2}{\epsilon_2}}
\end{eqnarray}
the previous equations are simplified as follows. For both cases (perpendicular and  parallel to plane of incidence), we obtain	
\begin{eqnarray}
\frac{E_t}{E_i}&=&\frac{2\cos i}{\cos i+\sqrt{1-(\frac{\hat{n}_1}{\hat{n}_2})^2\sin^2 i}}\nonumber\\
&=&\frac{2\cos i}{\cos i+\cos j}\nonumber\\
& \equiv &   T(i,j) \label{eqn:T(i,j)}
\end{eqnarray}
where $j$ is outgoing angle of light and

\begin{eqnarray}
\frac{E_r}{E_i}&=&\frac{\cos i-\sqrt{1-(\frac{\hat{n}_1}{\hat{n}_2})^2\sin^2 i}}{\cos i+\sqrt{1-(\frac{\hat{n}_1}{\hat{n}_2})^2\sin^2 i}}\nonumber\\
&=&\frac{\cos i-\cos j}{\cos i+\cos j}\nonumber\\
& \equiv & R(i,j)  \label{eqn:R(i,j)}
\end{eqnarray}
Let us consider that the light enters from the first Riemann sheet $W_{1f}$ to the second Riemann sheet $W_{1s}$ through the branch cut (red (dahed) line in Fig. \ref{fig: w1}). In what follows, we call "{\it first branch cut}" for branch cut from $W_{1f}$ to $W_{1s}$ and "{\it second branch cut}" for branch cut from $W_{1s}$ to $W_{1f}$. At a first look at Eq. (\ref{eqn:R(i,j)}), it seems that there should be some reflections at the first branch cut even if we make use of impedance matching. (See red arrows in Fig. \ref{fig:reflection image}.) However, we will show that this reflection is completely canceled out by summing up all the reflections that take place at the second branch cut.

Next, let us consider another case that the light enters from $W_{1f}$ to $W_{1s}$ through the first branch cut (orange incident arrow in Fig. \ref{fig:reflection image}). This light ray goes around the second Riemann sheet (orange arrows in Fig. \ref{fig:reflection image}) and is reflected at the second branch cut. Then, it returns to the first branch cut and goes through the first branch cut again (see blue arrows in Fig. \ref{fig:reflection image}). In this case, the ratio is given by
\begin{eqnarray}
&&A_1=T(i,j)R(j,i)T(j,i)
\end{eqnarray}

Furthermore, the light rays can be reflected again at the first branch cut and go
and back two times between the first branch cut and the second branch cut in $W_{1s}$, and finally come back to $W_{1f}$ through the first branch cut. For the case that the light rays go and back two times, the ratio is
\begin{eqnarray}
&&A_2=T(i,j)R^3(j,i)T(j,i)
\end{eqnarray}

More generally, for going back $n$ times, the ratio is
\begin{eqnarray}
&&A_n=T(i,j)R^{2n-1}(j,i)T(j,i)
\end{eqnarray}

The total sum of all contribution is
\begin{eqnarray}
A_{tot}&=&T(i,j)\{R(j,i)+R^3(j,i)+R^5(j,i)+\dots\}T(j,i)\nonumber\\
&=&T(i,j)\{\sum_{i=1}^{\infty}R^{2n-1}(j,i)\}T(j,i)
\end{eqnarray}

We insert Eq. (\ref{eqn:T(i,j)}) and Eq. (\ref{eqn:R(i,j)}) into the previous equation, and
  the total sum of the contributions can be computed as follows.
\begin{eqnarray}
A_{tot}&=&T(i,j)\frac{R(j,i)}{1-R^2(j,i)}T(j,i)\nonumber\\
&=&\frac{\cos j-\cos i}{\cos i+\cos j}
\end{eqnarray}

 Unexpectedly, the sum of all contributions cancel the original reflection $R(i,j)$ of Eq. (\ref{eqn:R(i,j)}). Therefore, there is no reflection at the branch cut at all.
 The perfect cancelation of reflection phenomena is surprising itself and can be seen
as a consequence of the proposed plus-minus construction.

\subsubsection{Discontinuity of refractive index at the boundary between positive and negative refraction indices}

Next, we discuss about the discontinuity of the boundary between positive and negative refraction indices. (See blue (dotted) line in Fig. \ref{fig: w1})

If we set $\epsilon_1=-\epsilon_2$ and $\mu_1=-\mu_2$ at the boundary, we obtain\begin{eqnarray}
\hat{n}_1=-\hat{n}_2
\end{eqnarray}
In this case, by considering Eq. (\ref{eqn:Reflection 1}) and Eq. (\ref{eqn:Reflection 2}) (or Eq. (\ref{eqn:R(i,j)})), we see that there is no reflection at all for all incident angle and polarization at the boundary between positive and negative refractive materials.
\begin{eqnarray}
&&\frac{E_r}{E_i}=0
\end{eqnarray}

As a conclusion, there is no reflection for both the boundary at branch cut (red (dashed) line in Fig. \ref{fig: w1}) and the boundaries between positive and negative refractive materials (blue (dotted) line in Fig. \ref{fig: w1}). Therefore, this construction is perfect in theory, with no time delay and complete absence of reflection.

It is worth noticing that the perfect invisibility effect (i.e., no time delay and no reflection) is independent of the precise refractive index profile. In order to create a perfect cloaking, it is enough to enclose the light trajectory in $W_{1s}$ by a proper symmetric positive and negative refractive index (plus-minus construction). Therefore, the current proposed construction is just one example because many other plus-minus constructions that lead to perfect invisibility are possible using the same principles.

\section{Summary}
We have presented a new design of perfect cloaking devices. The design involves a new method as well as a construction based on negative refraction material.
The novel methodology shows that we do not need to use the equation of motion for light rays. The trajectory of light is obtained easily by conformal mapping rotation, which simplifies the theoretical construction of the model. Second, the negative refraction material is crucial to achieve perfect invisibility. The results not only indicate that there is no time delay
but also show that the reflection phenomena is completely canceled out thanks to the plus-minus construction.

The finding that reflection is completely suppressed suggests that the plus-minus construction may be potentially applied to various areas and technologies outside of purely invisibility oriented devices. In our view the fundamental roots of the complete cancelation of light reflection are worthy of further exploration. The absolute control of reflection phenomena promises to bring significant technical improvements in diverse areas, encompassing electrical engineering, telecommunications, optoelectronics and microelectronics. The practical possibility of guiding electromagnetic fields in absolute absence of reflection phenomena has relevance to the design and implementation of many emerging applications such as lenses, medical imaging, microwave passive devices, radar/defense systems and wireless communications, from cell phones to Global Positioning Systems.

Impedance matching is the traditional  method to reduce the reflection for various devices. Here, we have demonstrated that a design based on the plus-minus construction can completely eliminate the reflection that occurs at boundaries between two materials having different indices of refraction, even though both indices are positive. Taken together, the invented method suggests a new way of reducing the reflection, beyond the classical impedance matching, that may lead to improve the efficiency of a large variety of optical and electromagnetic devices.

\begin{acknowledgments}
T.O. and J.C.N. gratefully acknowledge the funding support of a Grant-in-Aid for Scientific Research (C) from MEXT, Japan.  We thank Prof. Leonhardt and Prof. Jiang
for fruitful discussions.
\end{acknowledgments}






\end{document}